\documentclass[prd,amsmath,amssymb,superscriptaddress,twocolumn,10pt]{revtex4}%superscriptaddress,showpacs,

\pdfoutput=1

\usepackage{graphicx}
\usepackage{dcolumn}
\usepackage{bm}
\usepackage{amssymb}
\usepackage{latexsym}
\usepackage{booktabs}
\usepackage{amsmath}
\usepackage{multirow}
\usepackage[colorlinks=true, linkcolor=red, citecolor=blue]{hyperref}

\newcommand{\be}{\begin{equation}}
\newcommand{\ee}{\end{equation}}
\newcommand{\bq}{\begin{eqnarray}}
\newcommand{\eq}{\end{eqnarray}}

\usepackage[usenames,dvipsnames]{xcolor}

\begin{document}

\title{A new cosmological probe using super-massive black hole shadows}

\author{Jing-Zhao Qi}
\affiliation{Department of Physics, College of Sciences, Northeastern
University, Shenyang 110819, China} 
\author{Xin Zhang\footnote{Corresponding author}}
\email{zhangxin@mail.neu.edu.cn}
\affiliation{Department of Physics, College of Sciences, Northeastern
University, Shenyang 110819, China} 
\affiliation{Ministry of Education's Key Laboratory of Data Analytics and Optimization
for Smart Industry, Northeastern University, Shenyang 110819, China}
\affiliation{Center for High Energy Physics, Peking University, Beijing 100080, China}
\affiliation{Center for Gravitation and Cosmology, Yangzhou University, Yangzhou 225009, China}

\begin{abstract}
We study the prospects of using the low-redshift and high-redshift black hole shadows as new cosmological
standard rulers for measuring cosmological parameters. We show that, using the low-redshift observation of the black hole shadow of M87$^\star$, the Hubble constant can be independently determined with a precision of about 13\% as $H_0=70\pm 9$ km s$^{-1}$ Mpc$^{-1}$. The high-redshift observations of super-massive black hole shadows may accurately determine a combination of parameters $H_0$ and ${\Omega_{\rm m}}$, and we show by a simple simulation that combining them with the type Ia supernovae observations would give precise measurements of the cosmological parameters.

\end{abstract}
%\pacs{95.36.+x, 98.80.Es, 98.80.-k}
\maketitle

\section{Introduction}
The first ever image of the super-massive black hole M87$^\star$ was recently captured by the Event Horizon Telescope (EHT) \cite{Akiyama:2019eap}, which opened a new era of studying black hole physics. It was shown in Ref.~\cite{Tsupko:2019pzg} that the shadow of super-massive black hole can be used as a standard ruler cosmology. 

In Ref.~\cite{Bisnovatyi-Kogan:2018vxl}, it was reported that the angular size of a black hole shadow in an expanding universe can be calculated with a high accuracy as $\alpha_{\rm sh}(z)=3\sqrt{3}m/D_A(z)$, where $\alpha_{\rm sh}$ is the angular radius of the black hole shadow, $z$ is the redshift of the expanding universe, $m=GM/c^2$ is the mass parameter of a black hole with mass $M$, and $D_A$ is the angular diameter distance in cosmology. Thus, if $\alpha_{\rm sh}$ can be measured by the very long baseline interferometry (VLBI) technology and $M$ can be independently measured by another astrophysical method, then the angular diameter distance $D_A$ can be obtained by the measurement of shadows of the super-massive black holes. 

As shown in Ref.~\cite{Tsupko:2019pzg}, the shadow of a super-massive black hole could be observed in the low-redshift range ($z\lesssim 0.1$) and in the high-redshift range ($z>$ a few). In the redshift range $z\lesssim 0.1$, the radius of the shadow of a super-massive black hole can be larger than 1 $\mu$as if its mass is above $10^9~M_\odot$, where $M_\odot$ if the solar mass. Using the low-redshift measurements of the shadows, combined with independent measurements of the masses of super-massive black holes, one can obtain the Hubble constant $H_0$ independently of the cosmic distance ladder. On the other hand, due to the fact that the angular diameter distance decreases for redshifts $z\gtrsim 1$ in a realistic cosmological model, the angular size of a shadow is expected to increase for high redshifts, and thus the shadow of a super-massive black hole might be observed. As an example, for $z\sim 10$, the shadow of a black hole with a mass comparable with  M87$^\star$ is only about one order of magnitude smaller than of M87$^\star$. Therefore, if the mass of a high-redshift super-massive black hole can also be measured by an astrophysical method, then its angular diameter distance can also be measured using its shadow. This potentially provides a new cosmological probe for the high-redshift range of the universe. 

The aim of this paper is the following: (1) We wish to estimate the accuracy of the measurement of the Hubble constant that can be achieved using the black hole shadow of M87$^\star$ as a probe. It is well-known that there is a great tension concerning the Hubble constant, at a significance of about 4.4$\sigma$, between the local measurement using a distance ladder \cite{Riess:2019cxk} and the cosmological fit using the Planck observation of the cosmic microwave background (CMB) anisotropy \cite{Aghanim:2018eyx}. The distance ladder gives a higher value of $H_0$, and the Planck observation favors a lower value of $H_0$, and it is presently not known which is correct. This is a great puzzle in contemporary cosmology, and it is often said that cosmology is at a crossroad. Obviously, solving this puzzle needs a fully independent measurement of $H_0$. The black hole shadow method can potentially provide such a measurement. (2) We wish to investigate the role that the black hole shadow method could play in estimating cosmological parameters. The new cosmological probe provided by the black hole shadow method could explore the expansion history of the universe in the high-redshift range, for which the current cosmological probes cannot provide any effective exploration. It is thus necessary to understand if the black hole shadow method could be developed into a truly useful cosmological probe and what role it could play in parameter estimation.

\begin{figure}
\centering 
\includegraphics[scale=0.4]{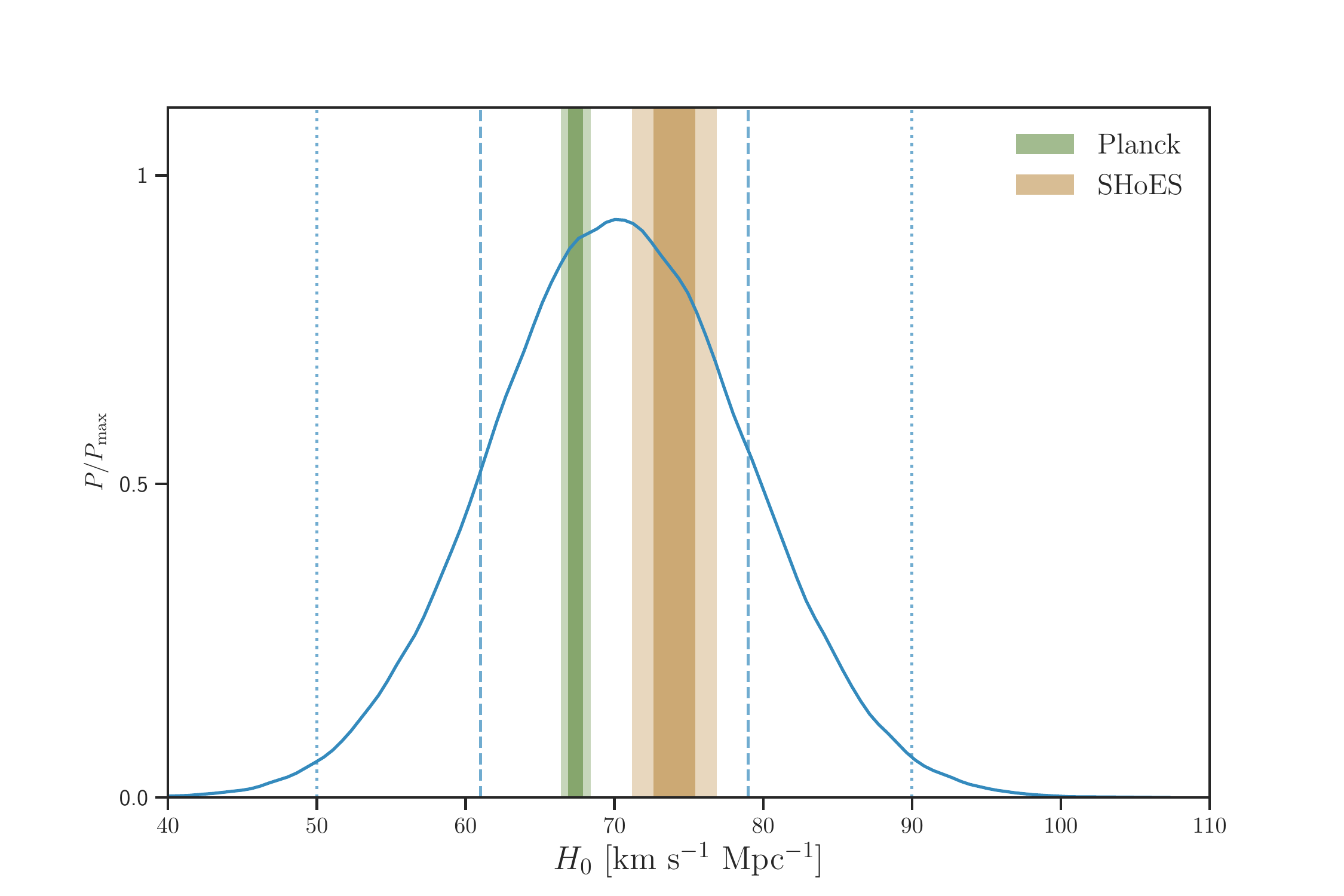}
\caption{The Hubble constant $H_0$ determined by the black hole shadow of M87$^\star$.  
One-dimensional posterior distribution of $H_{0}$ is shown by the blue curve, and the $68.3\%$ ($1\sigma$) and $95.4\%$ ($2\sigma$) confidence level intervals are indicated by dashed and dotted lines, respectively. The constraints on $H_0$ at $1\sigma$  (darker shading) and $2\sigma$ (lighter shading) from Planck 2018 \cite{Aghanim:2018eyx} and SHoES 2019 \cite{Riess:2019cxk} are also shown in green and orange, respectively.}\label{H0_fig} 
\end{figure}

\section{Measurement of the Hubble constant}

Let us first discuss the low-redshift case. The Hubble constant $H_0$ can be determined by the black hole shadow method, independent of the distance ladder and cosmology. At nearby distances, the Hubble constant $H_0$ can be directly measured by the Hubble law $v_{\rm H}=H_0 d$, where $v_{\rm H}$ is the velocity of the local ``Hubble flow'' of a source, and $d$ is the distance to the source. At such near distances, less than about 50 Mpc, all cosmological distance measurements are different only at the order of $v_{\rm H}/c$, where $c$ is the speed of light. Thus, in this case we approximately have $d=D_A$. 

It is well-known that the nearby massive elliptical galaxy M87 is the central galaxy of the Virgo cluster, which is the closest galaxy cluster to the Milky Way. The central radio source in M87, namely the black hole M87$^\star$, has a mass that can be measured by the stellar-dynamics observation, with a recent result obtained by Gebhardt et al. \cite{Gebhardt:2011yw} of $M=(6.6\pm 0.4)\times 10^9~M_{\odot}$. The black hole shadow of M87$^\star$ was measured by EHT \cite{Akiyama:2019eap}, with the angular radius of $\alpha_{\rm sh}=21.0\pm 1.5$ $\mu$as. Therefore, using the relationship of the standard ruler provided by the black hole shadow method $D_A=3\sqrt{3}m/\alpha_{\rm sh}$, one can immediately obtain the distance to M87$^\star$.

It is necessary to point out that the existing measurements of the black hole mass of M87$^\star$ are not in agreement. Another method, based on the gas-dynamics observation, gives a different result. The latest gas-dynamics observation by Walsh et al. \cite{Walsh:2013uua} gave the result of $M=(3.5^{+0.9}_{-0.7})\times 10^9~M_{\odot}$. Thus, the two methods (stellar-dynamics and gas-dynamics modelings) give inconsistent results for the black hole mass, with a difference of almost a factor of 2 (see Ref.~\cite{Akiyama:2019eap} for a brief review). In addition, another serious problem is that both measurements were obtained with a presumption of the distance of M87. Therefore, they cannot in principle be directly used to infer the distance of the black hole shadow.

This predicament is expected to be solved by another dynamic approach like the maser observations \cite{Miyoshi:1995da,kuo2018estimating,darling2017detect}, which can determine the central black hole mass with a few percent precision, independently of the distance to the galaxy. The Megamaser Cosmology Project (MCP) \cite{Zhao:2018qww,Kuo:2010uy} has conducted maser observations of a number of galaxies with redshifts up to $z\sim 0.05$. MCP can also observe the rotation velocities of masers in the disk near the black hole and their accelerations. With such information, not only the black hole mass but also the linear size of the disk can be determined, which allows an independent measurement of the Hubble constant \cite{Gao:2015tqd, Reid:2012hm}.

In this work, we study the prospects of using super-massive black hole shadows as new cosmological probe. Currently, we have rather limited observational information, and the discussion of the determination of the Hubble constant using the black hole shadow of M87$^\star$ serves only to illustrate the potential of the method. We thus need to make some assumptions. In this work, we adopt the black hole mass of Gebhardt et al. \cite{Gebhardt:2011yw}, $M=(6.6\pm 0.4)\times 10^9~M_{\odot}$.

In addition, we also need to clarify the relation $\alpha_{\rm sh}(z)=3\sqrt{3}m/D_A(z)$ used in this work. The value of $3\sqrt{3}m$ is derived with the assumption of the Schwarzschild metric. There are two factors that may influence the size of the black hole shadow. (i) The complicated accretion flow near the black hole might lead to a slightly larger size of the bright ring, depending on the emission profile \cite{Akiyama:2019eap, Tsupko:2019pzg}. (ii) An astrophysical black hole typically has a spin angular momentum, which can affect the shadow size at the level of $4\%$ \cite{Akiyama:2019eap, Takahashi:2004xh}. For M87$^\star$, the relation between its mass and the size of its shadow was determined by accurate numerical modeling. Therefore, the value of $3\sqrt{3}m$ can be safely used in our calculations. To characterize the effect of spin, we additionally consider an uncertainty of the shadow size at the level of $4\%$.

To determine the Hubble constant, one needs to obtain the Hubble flow velocity at the position of M87. We know that M87 is a part of the Virgo cluster, which has a center-of-mass recession velocity of $1283 ~ \rm{km~s^{-1}}$ (with the redshift $z=0.00428$ \cite{M87-redshift}). Usually, the typical peculiar velocity is about 10\% of the total recessional velocity at a nearby distance, and we adopt the value of the peculiar velocity of M87 of $v_{\rm p}=150$ km s$^{-1}$ \cite{lisker2018active}. The error of the peculiar velocity is rather difficult to estimate, and we assume a conservative 50\% error. From these values, we obtain the Hubble velocity $v_{\rm H}=1133\pm 75~ \mathrm{km~ s}^{-1}$. Once the distance and the Hubble velocity are determined, we can constrain the value of the Hubble constant, and we obtain the result of $H_0=70\pm 9$ km s$^{-1}$ Mpc$^{-1}$. This is a determination of the Hubble constant with a 13\% error. In Fig.~\ref{H0_fig}, we show the marginalized posterior distribution of our determination of $H_0$. 

The results of $H_0$ from Planck 2018 \cite{Riess:2019cxk} and SHoES 2019 \cite{Aghanim:2018eyx} are also shown (as green and orange bands) in Fig.~\ref{H0_fig}. It is clear that our determination of $H_0$ using the black hole shadow of M87$^\star$ cannot be used to arbitrate the Hubble constant tension, because currently we have only one data point and the uncertainty of determination of $H_0$ is large, around 13\%.  It could be expected that the measurement accuracies of the shadows and masses of super-massive black holes will be improved in the future. However, if we assume naively that the errors of single data points are similar, then the error of $H_0$ from the measurements of black hole shadow standard rulers would be $13\%/\sqrt{N}$, where $N$ is the number of measurements. Thus, about 40 data points would be needed to reduce the error of $H_0$ to about 2\%, comparable to the error of the current results.

Clearly, such a programme would be impossible in the near future due to the enormous challenges in observing black hole shadows and measuring their masses (see also the relevant discussion in Ref.~\cite{Tsupko:2019pzg}). Therefore, considering these challenges, the new probe based on the black hole shadows is presently premature.

We mention here the gravitational-wave standard siren measurement of the Hubble constant from the multi-messenger observation of the binary neutron star merger event (GW170817) \cite{GBM:2017lvd}, with the result of $H_0=70^{+12}_ {-8}$ km s$^{-1}$ Mpc$^{-1}$ \cite{Abbott:2017xzu}, which is also independent of the distance ladder. In Ref.~\cite{Chen:2017rfc}, it is reported that a 2\% $H_0$ measurement from standard sirens will be achieved within five years (with about 50 events to be observed by the Advanced LIGO-Virgo network). We expect that the black hole shadow standard ruler observations could also be greatly developed in the forthcoming years.

\section{Cosmological application of the high-redshift observations}
We wish to further discuss the high-redshift case. At large cosmological distances, it is possible to use the angular size of the black hole shadow as the standard ruler to probe the expansion history of the universe at very high redshifts ($z\sim 10$), which are extremely hard to explore with existing observations, provided that the mass of the super-massive black hole can be determined independently. As shown in Ref.~\cite{Tsupko:2019pzg}, to measure the angular size of a super-massive black hole in the high-redshift range, it is necessary to reach the angular resolution of about 0.1 $\mu$as. Although this is extremely challenging, it could be possible, as shown in Ref.~\cite{Tsupko:2019pzg}, once the VLBI technology in optical bands is developed as it can increase the resolution by orders of magnitude due to the shorter wavelengths employed. To determine the mass of super-massive black hole at high redshifts, one can consider using the reverberation mapping method that is another way to determine black hole mass dynamically, besides the standard methods based on stellar or gas dynamics for nearby black holes. At present, the uncertainties of mass determination of black holes are still large. However, as the observations and theoretical understanding improve, it is possible that the situation will change in the future and that the mass of super-massive black holes at high redshifts could be determined with higher precision. 

\begin{figure}
\centering 
\includegraphics[scale=0.45]{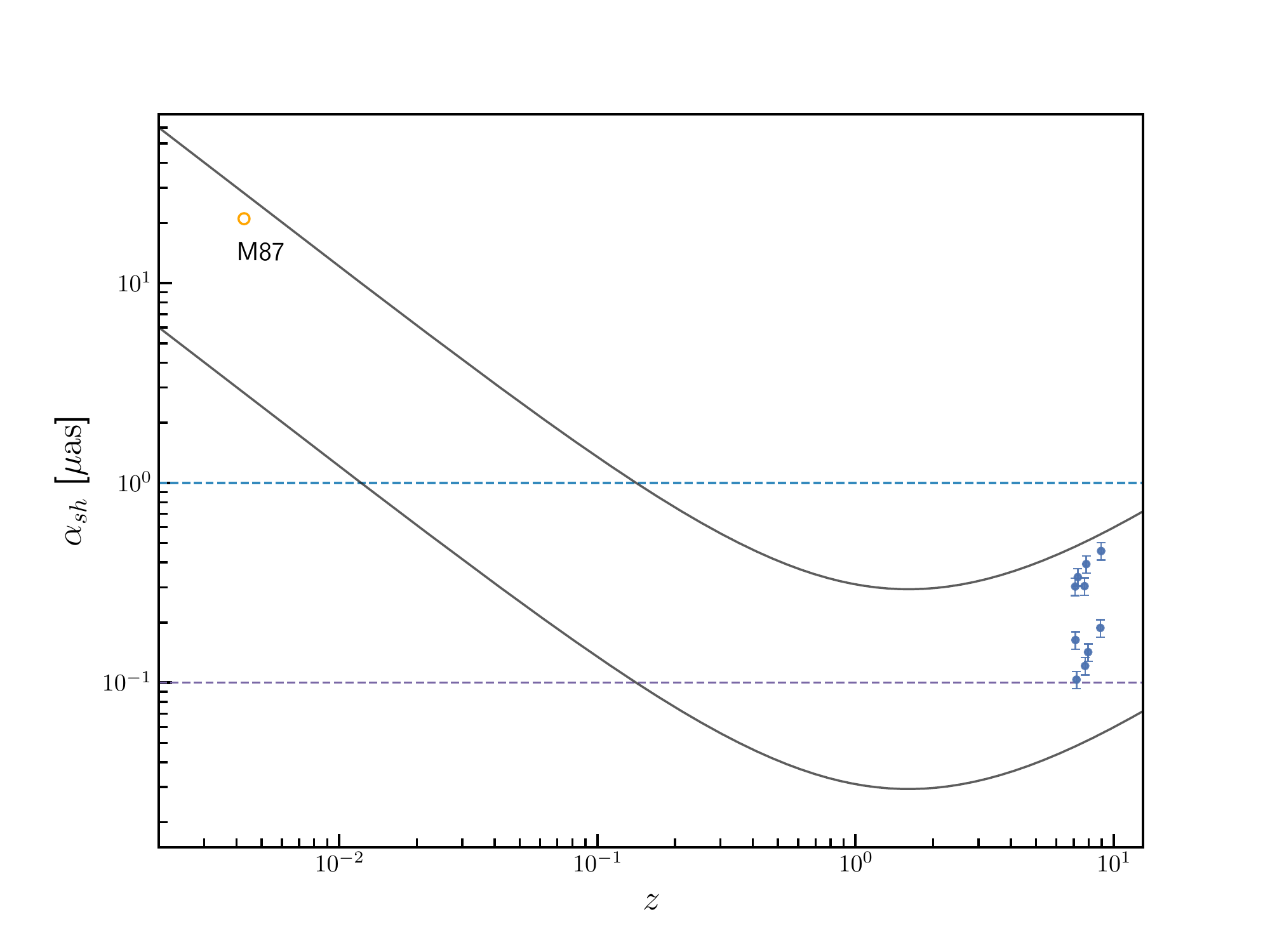}
\caption{Simulation data for the black hole shadows at high redshifts (shown as blue dots with error bars). Two black curves represent the angular radius $\alpha_{\rm{sh}}(z)$ of the super-massive black holes with a mass of $10^9$ (bottom) and $10^{10}~M_{\odot}$ (top). M87$^\star$ is shown as the orange circle.}\label{Ash_fig} 
\end{figure}

Hence, it is of great interest to understand what role the cosmological probe based on the black hole shadow method could play in the estimation of cosmological parameters. We use a simple simulation to investigate this issue. In the following, we use 10 simulated data points to perform this analysis. The simulated data are shown in Fig.~\ref{Ash_fig}. In the simulation, we employ the standard flat $\Lambda$ cold dark matter ($\Lambda$CDM) model (with $\Omega_{\rm m}=0.3$ and $H_0=70$ km s$^{-1}$ Mpc$^{-1}$) as the fiducial model. We assume that the resolution of the angular size measurement of the black hole shadows can reach 0.1 $\mu$as, and that the masses of black holes can be measured with an uncertainty of around 10\%. We simulate the data with a uniform distribution in the redshift interval $z\in [7,9]$ and in the mass interval $M\in [10^9, 10^{10}]~M_{\odot}$, as shown in Fig.~\ref{Ash_fig}. In the figure, the two black solid lines show the evolution of $\alpha_{\rm sh}(z)$ of a super-massive black hole with a mass of $10^9$ (bottom) and $10^{10}$ $M_{\odot}$ (top). M87$^\star$ (orange circle) is also shown in the figure.

\begin{figure}
\centering 
\includegraphics[scale=0.4]{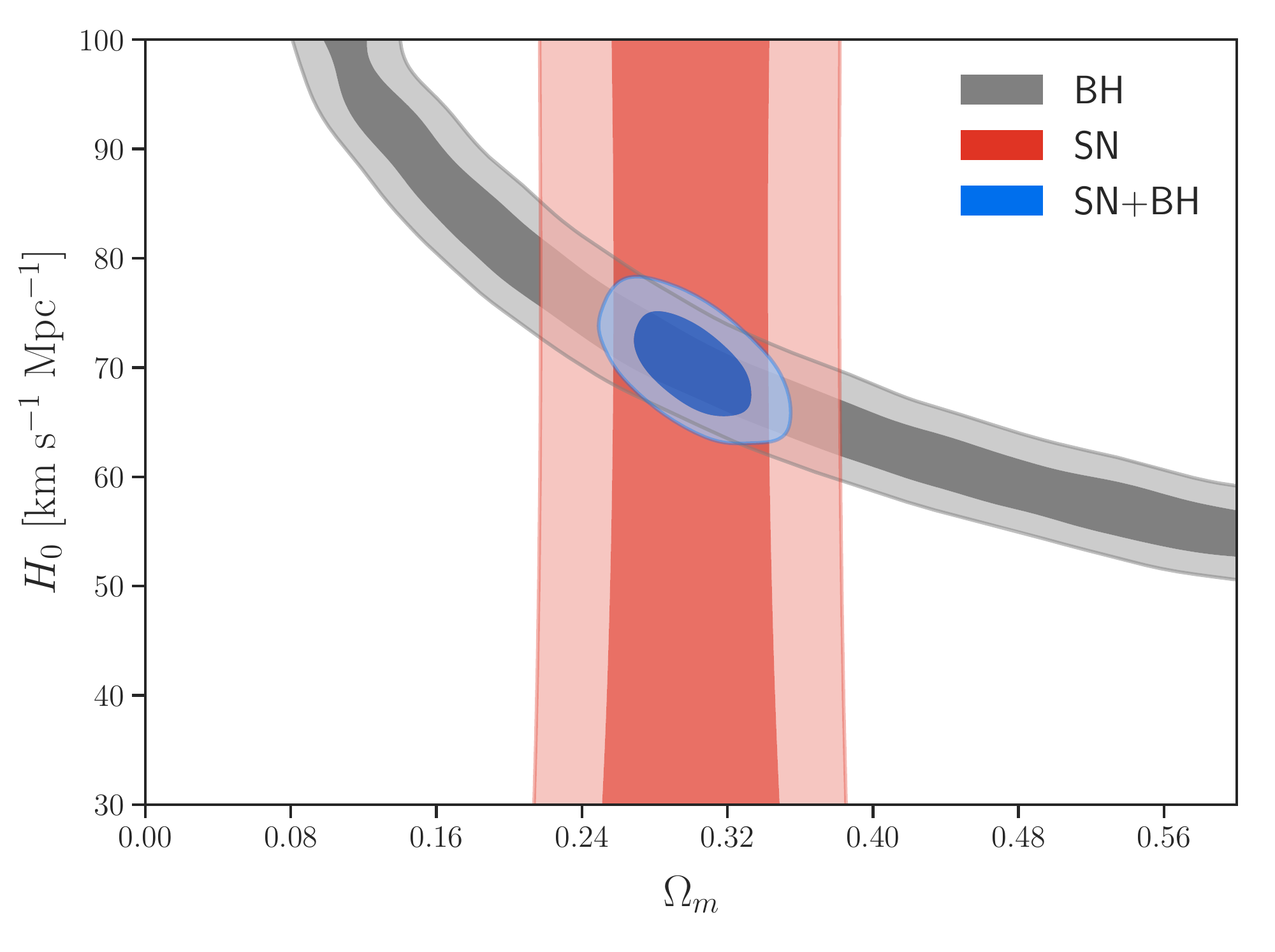}
\caption{The constraints ($68.3\%$ and $95.4\%$ confidence level) on the $\Lambda$CDM model obtained by using the simulated black hole shadow data (gray contours), the latest Pantheon compilation of type Ia supernova data (red contours), and the combination of the two data sets (blue contours).}\label{lcdm_fig} 
\end{figure}

We use the simulated data to constrain the $\Lambda$CDM model, shown in Fig.~\ref{lcdm_fig}. It can be clearly seen that, with 10 high-redshift black hole shadow data, neither $\Omega_{\rm m}$ nor $H_0$ can be tightly constrained. Only a combination of $\Omega_{\rm m}$ and $H_0$ can be accurately determined, exhibited as a grey narrow strip in the $\Omega_{\rm m}$--$H_0$ plane. 

Comparing the constraints on $\Omega_{\rm m}$ and $H_0$, we observe that the constraint on $\Omega_{\rm m}$ is much weaker. This is an indication that the constraints from the type Ia supernovae (SN) observation might provide a helpful complement to the black hole shadow measurements, because the SN observation can only tightly constrain $\Omega_{\rm m}$ but is unable to constrain $H_0$ without the calibration by the cosmic distance ladder. Therefore, we also use the latest Pantheon compilation \cite{Scolnic:2017caz} of the SN observation to constrain the $\Lambda$CDM model, with the result shown as a red band in the $\Omega_{\rm m}$--$H_0$ plane in Fig.~\ref{lcdm_fig}. We can clearly see that the parameter degeneracies generated by the black hole shadow and SN observations are thoroughly broken. The combined data sets give a rather tight constraint: $\Omega_{\rm m}=0.301\pm0.022$ and $H_0=70.3\pm 3.1$ km s$^{-1}$ Mpc$^{-1}$.

In fact, as discussed above, the low-redshift black hole shadow observations can be used to measure the Hubble constant independently of the cosmological models. Therefore, if the low-redshift black hole shadow observations are directly combined with the high-redshift observations, the new probe can determine the cosmological parameters independently without any other external methods. For this purpose, we simulated 10 low-redshift data sets and combined them with the high-redshift data to constrain the $\Lambda$CDM model. We find that the degeneracies between $\Omega_{\rm m}$ and $H_0$ can be broken, and that the joint constraint is $\Omega_{\rm m}=0.315^{+0.049}_{-0.067}$ and $H_0=70.0^{+2.9}_{-3.5}$ km s$^{-1}$ Mpc$^{-1}$.

\section{Conclusion}
We discussed in this paper the prospects of using the black hole shadows as cosmological standard rulers to measure the cosmological parameters. In the low-redshift range, the black hole shadows can be used to measure the Hubble constant independently of the distance ladder. As an example, using the present observation of the black hole shadow of M87$^\star$ combined with the measurement of its mass with the stellar-dynamics observation, we determined the Hubble constant as $H_0=70\pm 9$ km s$^{-1}$ Mpc$^{-1}$ with a precision of about 13 \%. This result cannot be used to arbitrate the Hubble constant tension. A naive estimate shows that to improve the precision of the measurement of the Hubble constant to 2\%, about 40 observations of the black hole shadows as standard rulers would be needed. It is also possible to use the angular size of the black hole shadow as a standard ruler to probe the expansion history of the universe at very high redshifts ($z\sim 10$), provided that the black hole mass can be determined independently. This is extremely challenging but not impossible in the future. It is thus of great interest to understand what role such a cosmological probe could play in estimating the cosmological parameters. We studied this case by using a simple simulation, and found that with only 10 simulated data points, a combination of $H_0$ and ${\Omega_{\rm m}}$ can be constrained (shown as a narrow strip in the ${\Omega_{\rm m}}$--$H_0$ plane). We further showed that the SN observations can be a useful complement to this cosmological probe, and that the combination with SN observations could fully break the parameter degeneracy, giving tight constraints on $H_0$ and ${\Omega_{\rm m}}$. 

It should be stressed that although the principle underlying the black hole shadow observation is very simple, its cosmological application is absolutely not. Although the black hole shadows have the potential to be new cosmological probes, due to the fact that their observation is very challenging, the method is still primitive and premature. We expect that the black hole shadow standard rulers will be developed into important cosmological probes in the future, significantly promoting the development of cosmological parameter measurements.

\begin{acknowledgments}
We are grateful to the anonymous referees for very useful comments helping us to significantly improve the paper. We would also like to thank Heng Yu for helpful discussions. This work was supported by the National Natural Science Foundation of China (Grants Nos.~11975072, 11835009, and 11690021), the National Program for Support of Top-Notch Young Professionals, and the Fundamental Research Funds for the Central Universities (Grant No. N180503014).

\end{acknowledgments}

\bibliography{BH_ref}

\end{document}